\documentclass[aps,twocolumn,showpacs,preprintnumbers,nofootinbib,prd,superscriptaddress,groupedaddress]{revtex4}
\pdfoutput=1
\usepackage{amssymb,graphicx}
\usepackage{amsmath}
\usepackage{multirow}
\usepackage{epsfig}
\usepackage[usenames]{color} 
\usepackage[export]{adjustbox}
\usepackage{mathtools}

\newcommand{\beqn}{\begin{eqnarray}}
\newcommand{\eeqn}{\end{eqnarray}}
\newcommand{\beq}{\begin{equation}}
\newcommand{\eeq}{\end{equation}}

\def\mphi{m_{\phi}}

\def\gammab{\hat{\gamma}}

\def\psib{\bar{\psi}}
\def\Phib{\overline{\Phi}}
\def\tg{\tilde{g}}

\def\mcR{\mathcal{R}^2}
\def\meff{m_{\textrm{eff}}}

\begin{document}

\title{Spontaneous tensorization from curvature coupling and beyond}
\author{Fethi M.\ Ramazano\u{g}lu}
\affiliation{Department of Physics, Ko\c{c} University, \\
Rumelifeneri Yolu, 34450 Sariyer, Istanbul, Turkey }
\date{\today}

\begin{abstract}
We generalize the recently introduced extended scalar-tensor-Gauss-Bonnet
(ESTGB) theories and their close relatives to include spontaneous growth of
nonscalar fields such as vectors. This is analogous to the program that developed spontaneous
tensorization from the original spontaneous scalarization theory of Damour and
Esposito-Far{\`e}se (DEF). The new larger family of theories conserves the appeal of
the DEF theory in terms of conforming to weak-field tests and also
providing large signals in strong gravity. Moreover, they provide a much richer
phenomenology including spontaneous tensorization of black holes as in ESTGB.
These theories, together with other possible future extensions that we discuss, testify to
the ubiquity of spontaneous growth in gravity. We also note that
theories with derivative coupling require special attention since  they 
can lead to potentially problematic higher derivatives in the equations
of motion.
\end{abstract}
\maketitle

\section{Introduction}
The age of gravitational waves has provided new opportunities to
test gravity in the strong and dynamical field regime, which in turn
has been driving many efforts to understand possible deviations
from general relativity (GR)~\cite{0264-9381-32-24-243001,Barack:2018yly}.
One of the leading avenues in this 
area has been the investigation of spontaneous scalarization.
This phenomenon was originally proposed 
by Damour and Esposito-Far{\`e}se (DEF) in scalar-tensor theories,
where the matter coupling of the scalar leads to an effective scalar mass term with
the ``wrong'' sign~\cite{PhysRevLett.70.2220}. The resulting tachyonic instability
of the GR solution in the presence of matter causes astrophysical
neutron stars to have stable scalar clouds around them.

The main appeal of spontaneous scalarization is twofold. First, the scalar
dies off fast away from the star, and satisfies the weak-field tests that
strongly disfavor other variants of scalar-tensor theories~\cite{Will:2001LR}.
Second, deviations from GR are large near a neutron star, and hence are
easier to detect with gravitational wave observations which currently have 
relatively low precision~\cite{LIGOScientific:2018mvr}. These factors
have inspired the investigation of the spontaneous scalarization phenomena
in theories beyond their original context. A recent development in this
regard is the extended scalar-tensor-Gauss-Bonnet (ESTGB)
theories where the scalar couples to the Gauss-Bonnet term rather than
matter~\cite{Silva:2017uqg,Doneva:2017bvd,Antoniou:2017acq}.
The mechanism works similarly, where the tachyonic instability
is incited by curvature rather than matter, and hence can lead to scalarized
black holes as well as neutron stars. The idea can also be extended to
other higher-curvature terms, or to more generic terms such as an
Einstein-Maxwell-scalar action in a relatively straightforward
manner~\cite{Herdeiro:2018wub}.

A second program to generalize spontaneous scalarization has been
undertaken in parallel to ESTGB theories, where there is still matter
coupling as in the DEF theory, but the scalar is replaced by some other
field, e.g. a vector, or the nature of the coupling is modified to include derivative
terms. The former has led to theories of spontaneous vectorization~\cite{Ramazanoglu:2017xbl},
and the latter to spontaneous scalarization based on ghostlike instabilities rather than
tachyons~\cite{Ramazanoglu:2017yun}. These general phenomena
of spontaneous growth are categorized under the umbrella term ``spontaneous tensorization.''

The main theme of the current study is to combine the two aforementioned
avenues to generalize spontaneous scalarization. We consider fields that couple
to a general action term such as the Gauss-Bonnet term (instead of just matter).
However,  we allow the fields to be nonscalars, e.g. vectors, or we allow derivative couplings. Such
fields can also go through spontaneous growth with suitable coupling functions
which lead to a further generalization of spontaneous scalarization we
call ``extended spontaneous tensorization.''
These novel theories preserve the appeal of the DEF theory in
weak and strong gravity regimes while providing a richer
phenomenology. However, they may also feature higher-derivative
terms in the field equations in cases with derivative couplings,
which does not occur for derivative coupling to matter.
We will employ $G=c=1$
throughout the paper.

\section{Spontaneous scalarization in extended scalar-tensor theories}\label{sec:scalarization}
Let us first review the known generalizations of spontaneous scalarization
starting with ESTGB theory
for which the coupling of the Gauss-Bonnet term in the action depends on a dynamical scalar
field~\cite{Silva:2017uqg,Doneva:2017bvd,Antoniou:2017acq}\footnote{
This idea was proposed by three different groups, but our discussion is closer
to Refs.~\cite{Silva:2017uqg,Doneva:2017bvd} which 
emphasized the role of instabilities and spontaneous scalarization.}
\begin{align}\label{egb_action}
\frac{1}{16\pi} \int d^4x \sqrt{-g} [R-2\nabla_\mu\phi \nabla^\mu\phi
+\lambda^2 f(\phi) \mcR]\ ,
\end{align}
where $\mcR = R^2-4R_{\mu\nu}R^{\mu\nu}+R_{\mu\nu\rho\sigma}R^{\mu\nu\rho\sigma}$
is the Gauss-Bonnet invariant, $\phi$ is a dynamical scalar field with
curvature coupling $f(\phi)$, and $\lambda$ is a constant whose
dimension ensures that $f(\phi)$ is dimensionless.
 The most relevant equation of motion for our discussion is that
of the scalar
\begin{align}\label{egb_eom0}
\Box \phi = -\frac{\lambda^2}{4} f'(\phi) \mcR \ .
\end{align}
We will use $'$ to denote the derivative of any function with respect to
its argument, in this case $\phi$.

Assume that the first derivative of $f$ vanishes at $\phi=0$, and consider the linearization
of Eq.~\ref{egb_eom0}
\begin{align}\label{egb_eom}
\Box \phi \approx -\frac{\lambda^2}{4} \widehat{f''(\phi)} \mcR \phi  \equiv \meff^2\ \phi \ ,
\end{align}
where an overhat denotes the value of a function when its argument
(in this case $\phi$) is $0$.
This is the wave equation for a scalar field with effective mass $\meff$, even
though we have a massless scalar in the action. The case of interest to us is
when $f''(\phi)>0$ and $\meff$ is imaginary, i.e. $\phi$
is a tachyon. A Fourier mode $e^{i(\vec{k} \cdot \vec{x} - \omega t)}$
with $\omega \sim \pm \sqrt{k^2 +\meff^2}$ would undergo exponential 
growth rather than oscillations for small $k$. In other words, there is an infrared instability
around $\phi=0$, but depending on the form of $f(\phi)$, nonlinear terms
can eventually stop this growth leading to a stable scalar field
solution~\cite{Blazquez-Salcedo:2018jnn,Minamitsuji:2018xde,Silva:2018qhn}.
Such solutions were explicitly constructed for spherically symmetric black holes
and neutron stars, circumventing no-hair results~\cite{Silva:2017uqg,Doneva:2017bvd,Antoniou:2017acq}.

A standard function that provides all of these features is $f(\phi)=1-e^{\beta \phi^2/2}$
for some negative constant $\beta$~\cite{Doneva:2017bvd}, but any function with similar
parabolic behavior around $\phi=0$ and decaying derivatives at large scalar values
provides qualitatively similar results. The decay diminishes the effective mass term with
growing scalar field values, preventing a runaway instability.

One can immediately see that the specific form of the Gauss-Bonnet term
does not play a specific role in Eqs.~\ref{egb_action}-~\ref{egb_eom}, so 
replacing it with another nonvanishing action term still leads to spontaneous scalarization.
The action for the most general such theory is
\begin{align}\label{generic_scalar_action}
\frac{1}{16\pi} \int d^4x \sqrt{-g} [R-2\nabla_\mu\phi \nabla^\mu\phi
+\bar{\lambda}^2 f(\phi) \mathcal{L}_\chi[\chi, g_{\mu\nu}]]\ ,
\end{align}
with linearized scalar equation of motion
\begin{align}\label{generic_scalar_eom}
\Box \phi \approx -\frac{\bar{\lambda}^2}{4}\mathcal{L}_\chi  \widehat{f''(\phi)}\ \phi \equiv \meff^2\ \phi \ ,
\end{align}
if $\widehat{f'(\phi)}=0$. $\mathcal{L}_\chi$ can be a purely metric term as in 
Gauss-Bonnet or Chern-Simons gravity~\cite{Gao:2018acg}, or depend on another field
$\chi$ such as the Maxwell action of a vector field 
$\mathcal{L}_\chi = F_{\mu\nu}F^{\mu\nu}$~\cite{Herdeiro:2018wub}.
We will call this family of theories with generic $\mathcal{L}_\chi$
``extended spontaneous scalarization theories''.

Hairy compact object solutions are not a complete novelty, and the case of neutron stars
has been known since the scalar-tensor theories of Damour and Esposito-Far\`{e}se~\cite{PhysRevLett.70.2220}.
Consider the action
\begin{align}\label{def_action}
\frac{1}{16\pi} \int &d^4x \sqrt{-g} [R-2\nabla_\mu\phi \nabla^\mu\phi] \nonumber \\
+&S_m[\psi_m,A^2(\phi) g_{\mu\nu}]
\end{align}
where $S_m$ is the action for any matter field $\psi_m$, with
the peculiarity that matter does not  couple to $g_{\mu\nu}$
(minimal coupling), but to a conformally scaled metric
$\tg_{\mu\nu} = A^2g_{\mu\nu}$. This gives the scalar equation of motion
\begin{align}\label{def_eom}
\Box \phi = -8\pi\tilde{T}A^3 \widehat{\frac{dA}{d(\phi^2)}}\ \phi \ 
\end{align}
where $\tilde{T}$ is the trace of the matter stress-energy tensor
with respect to $\tg_{\mu\nu}$. The right-hand side behaves exactly as
an effective mass term for appropriate couplings such as $A=e^{\beta \phi^2/2}$
with negative $\beta$. It has been known that neutron stars scalarize in this theory in
a natural part of the parameter space. This choice of $A$ also
ensures that the instability gets weaker as the scalar field grows, leading to 
stable scalar clouds around neutron stars.

The extended spontaneous scalarization in Eq.~\ref{generic_scalar_action}
uses the same underlying mechanism as Eq.~\ref{def_action}: effective
mass generation through coupling to other fields. Despite the similarity
between the two, there is much richer phenomenology in the extended case.
For example, a scalar coupling to higher-derivative terms allows hairy black
holes whereas spontaneous scalarization in Eq.~\ref{def_eom} only scalarizes
matter fields.

As a last note, we should add that spontaneous scalarization can occur for
self-interacting scalar fields as well. The simplest case would be a massive
scalar $\nabla_\mu\phi \nabla^\mu\phi \to \nabla_\mu\phi \nabla^\mu\phi +m_\phi^2 \phi^2$.
The intrinsic mass term $m_\phi$ suppresses spontaneous growth
through its positive contribution to $\meff^2$, but one can still obtain
imaginary effective mass for similar choices of
$f(\phi)$. Despite this negative aspect,
a massive scalar dies off much faster in the far field, providing better
agreement with observations. This has been valuable for the DEF theory~\cite{Ramazanoglu:2016kul},
and might be useful in extended spontaneous scalarization theories as well.
For this reason, we will consider the more general massive fields in the 
following discussion.

\section{Extended vector-tensor theories}\label{sec:vectorization}
We have seen that the term that the scalar field couples to is not
crucial for spontaneous growth. A close examination shows that
the type of field that grows spontaneously is not essential either.
The latter idea has already been applied to the DEF theory where
one replaces the scalar in Eq.~\ref{def_action} with a vector
field $X_\mu$, 
\begin{align}\label{vt_action}
\frac{1}{16\pi} \int &d^4x \sqrt{-g}\ [R-F_{\mu\nu}F^{\mu\nu}-2m_X^2 X_\mu X^\mu] \nonumber \\
+S_m&[\psi,A_X^2(x) g_{\mu\nu}] 
\end{align}
where $F_{\mu\nu} = \nabla_\mu X_\nu -\nabla_\nu X_\mu$ and
$x =g^{\mu\nu}X_\mu X_\nu $. We use a massive vector (Proca) field
as we explained at the end of the previous section.
The vector field equation of motion is 
\begin{align} \label{eom_vt}
\nabla_\rho F^{\rho \mu} = (-8\pi A_X^4 \Lambda \tilde{T} +m_X^2) X^\mu \ ,
\end{align}
and the right-hand side behaves as an effective mass term in the Proca equation.
An appropriate choice of $A_X$ such as $e^{\beta_X X_\mu X^\mu}$ renders the left-hand side
negative, and leads to
spontaneous growth of $X_\mu$. This phenomenon is called ``spontaneous
vectorization''~\cite{Ramazanoglu:2017xbl}. The term that covers all
spontaneously growing fields is ``spontaneous tensorization''. 

The two ideas that generalize spontaneous scalarization in scalar-tensor theories can
be combined in an ``extended vector-tensor-Gauss-Bonnet (EVTGB) theory''
\begin{align}\label{egbvt_action}
\frac{1}{16\pi} \int d^4x \sqrt{-g}\ [&R-F_{\mu\nu}F^{\mu\nu}-2m_X^2 X_\mu X^\mu \nonumber \\
+&\lambda_X^2 f_X(x) \mcR] \ ,
\end{align}
with the vector equation of motion
\begin{align}\label{egbvt_eom}
\nabla_\rho F^{\rho \mu} = \left(-\frac{\lambda_X^2}{2} \mcR f_X'(x)
+m_X^2\right) X^\mu \ .
\end{align}
An appropriate choice of $f_X$ such as $f_X=1-e^{\beta_X X_\mu X^\mu/2}$
would give $X_\mu$ an imaginary effective mass,
leading to vectorized black holes and neutron stars. This can be immediately
generalized to
\begin{align}\label{generic_vector_action}
\frac{1}{16\pi} \int d^4x \sqrt{-g}\ \big[ &R-F_{\mu\nu}F^{\mu\nu}-2m_X^2 X_\mu X^\mu \nonumber \\
+& \bar{\lambda}_X^2 f_X(x) \mathcal{L}_\chi[\chi, g_{\mu\nu}] \big]\ ,
\end{align}
just as in the case of scalar fields in Eq.~\ref{generic_scalar_action}, with the equation of motion
\begin{align}\label{generic_vector_eom}
\nabla_\rho F^{\rho \mu} = \left( -\frac{\bar{\lambda}_X^2}{2} \mathcal{L}_\chi  f_X'(x)
+m_X^2 \right) X^\mu \ .
\end{align}
We will call this mechanism ``extended spontaneous vectorization.''
Similarly, the general name that covers all fields is ``extended spontaneous tensorization''.
We note that we do not claim that $X_\mu$ is any known part of the
Standard Model, in parallel with the case of $\phi$ in spontaneous
scalarization which is understood as a hitherto unobserved fundamental field.
\section{Extended spontaneous tensorization with ghostlike instabilities}\label{sec:ghost}
Extension of spontaneous growth from scalars to vectors is relatively straightforward, i.e.,
One changes the canonical scalar field action for the canonical vector field action,
and chooses a coupling function $f$ that provides an imaginary effective mass. 
A vector is no more special than a scalar, and hence spontaneous growth seems
to be extendable to other fields as well. However, we will first discuss
another path to generalize spontaneous scalarization: using a ghost
instead of a tachyon to incite an instability. This idea is again analogous to existing
ideas in gravity with nonminimal matter couplings (similar to the DEF theory),
but we will see that adapting 
such mechanisms to curvature couplings is not straightforward, and can
introduce new challenges. 

\subsection{Ghosts in spontaneous growth}
Changing the field that spontaneously scalarizes is not the only way to generalize
spontaneous growth beyond extended scalar-tensor theories. The growth
needs an instability, but this does not have to be a tachyon which has been the
exclusive case so far in our discussion. In principle, a ghostlike behavior can provide
the instability as well. Let us demonstrate this on a modification
of the massive DEF theory
\begin{align}\label{ghost_action}
\frac{1}{16\pi} \int &d^4x \sqrt{-g} [R-2\nabla_\mu\phi \nabla^\mu\phi -2\mphi \phi^2]
\nonumber \\
+& S_m[\psi,A_\partial^2(K) g_{\mu\nu}] \ , \ K=g^{\mu\nu} \partial_\mu \phi \partial_\nu \phi \ .
\end{align}
with the equation of motion
\begin{align}\label{ghost_eom}
\nabla_\mu \left[(-8\pi\tilde{T}A_\partial^3 A_\partial'+1) \nabla^\mu \phi\right] = m_\phi^2 \phi \ .
\end{align}

A function of the form $A_\partial = e^{\beta_\partial K/2}$ with $\beta_\partial <0$
would reverse the coefficient of the principal part of the partial differential equation
on the left, which can also be seen in the linearized equation around $\phi=0$
\begin{align}\label{ghost_eom}
(-4\pi\tilde{T}\beta_\partial+1) \Box \phi = m_\phi^2 \phi 
+4\pi \beta_\partial \nabla_\mu \tilde{T} \nabla^\mu \phi \ .
\end{align}
 A scalar with the
``wrong'' kinetic term is called a ghost, and it also causes exponential growth.
This phenomena is called
``ghost-based spontaneous scalarization''~\cite{Ramazanoglu:2017yun}. 
A physically relevant scalarization process would require the eventual suppression
of the growth due to nonlinear terms and the stability of the final scalarized
solution. The exponential form of $A_\partial$ suggests such a suppression,
but the stability of these solutions is not yet known. 
The coefficient of $\Box \phi$ is positive in the far field, and hence it has to vanish
at some point near the scalarization region. Such differential
equations can lead to problematic behavior, which requires detailed mathematical
analysis~\cite{Ramazanoglu:2017yun}.

This method of scalarizing neutron stars can also be adapted to more general
gravity theories just like the extended spontaneous scalarization theories in
(Eq.~\ref{generic_scalar_action}).
Namely, the action
\begin{align}\label{generic_scalar_action_ghost}
&\frac{1}{16\pi} \int d^4x \sqrt{-g} [R-2\nabla_\mu\phi \nabla^\mu\phi -2m_\phi^2 \phi^2]
\nonumber \\
+&\frac{1}{16\pi} \int d^4x \sqrt{-g}\ \bar{\lambda}_\partial^2  f_\partial(K) \mathcal{L}_\chi[\chi, g_{\mu\nu}]]\ ,
\end{align}
with the equation of motion
\begin{align}\label{generic_scalar_eom_ghost}
\nabla_\mu\left[ ( -\frac{\bar{\lambda}_\partial^2 }{2}\mathcal{L}_\chi f_\partial'(K) +1)\ \nabla^\mu \phi \right] = m_\phi^2 \phi 
\end{align}
would also have a ghostlike instability and spontaneous growth for a
choice such as $f_\partial=1-e^{\beta_\partial K}$ with appropriate $\beta_\partial$.

Despite the similarities between the original ghost-based spontaneous
scalarization through matter coupling (Eq.~\ref{ghost_eom}) and
through generic coupling (Eq.~\ref{generic_scalar_eom_ghost}), we
need to take additional care with the latter. In Eq.~\ref{ghost_eom}, the additional
terms that arise from the nonminimal coupling introduce at most
second-derivative terms into the equations of motion. This is still the case for the derivatives of
$\phi$ in Eq.~\ref{generic_scalar_eom_ghost}; however, we also have
$\nabla_\mu \mathcal{L}_\chi$. This term again brings at most second
derivatives if $\mathcal{L}_\chi$ is the Maxwell Lagrangian as in the
Einstein-Maxwell-scalar theories. On the other hand, if we couple
to the Gauss-Bonnet term, we now have $\nabla_\mu \mcR$ appearing in
the equation of motion, which has three derivatives
acting on the dynamical field $g_{\mu\nu}$. We have not closely examined
the metric equation of motion (the modified Einstein equation) in our study so far,
since it did not play a crucial role in the onset of the instability.
However, it also contains third derivatives, this time of the scalar
field.\footnote{Roughly, the variation
of the Gauss-Bonnet term with respect to the metric moves its two derivatives
to its coefficient $f_\partial$ through partial integration,
resulting in three derivatives of $\phi$.}

Aside from changing the
principal part of the equations of motion, which is a major modification,
more than two time derivatives are potential sources of unphysical behavior
through Ostrogradsky's theorem~\cite{Woodard:2015}. This is one of the main
reasons that makes the scalar coupling (without derivatives)
to the Gauss-Bonnet term very important: it is the only higher-curvature
term that provides a second-order partial differential equation
as the equation of motion. This aspect
of the theory is apparently lost once the coupling contains derivatives of the scalar.
In fact, any derivative term has the same effect as we will shortly see.

Extra derivatives mean that extended spontaneous tensorization with
derivative coupling should be
analyzed with more care than the direct scalar or vector field coupling,
but it can still be physically relevant.
First, theories with more than two derivatives in their equation
of motion can be viable due to symmetries which allow the expression
of higher derivatives in terms of lower ones, which ultimately renders them
second order, such as $f(R)$ gravity theories and certain Beyond-Horndeski
theories~\cite{Sotiriou:2008rp,Zumalacarregui:2013pma,Ramazanoglu:2019tyi}.
We could not identify such symmetries for Eq.~\ref{generic_scalar_eom_ghost},
but they can be nontrivial to identify. Second, even though the Gauss-Bonnet
term is unique in providing second-order equations in four dimensions,
there have been countless studies on other higher curvature terms~\cite{Barack:2018yly},
including theories with extended spontaneous scalarization~\cite{Gao:2018acg}.
Such theories are investigated using the ``effective field theory'' approach
where they are thought to be low-energy approximations to an
unknown well-behaved theory~\cite{Barack:2018yly}. Hence, the
higher-derivative terms are treated as perturbations to understand the well-behaved
theory, even though they form the principal part of the partial differential
equations when taken at face value. 
Equations of motion arising from Eq.~\ref{generic_scalar_action_ghost}
can also be treated in this manner.

We can consider other fields with
derivative couplings as well. For example, one can consider a theory of
vector derivative coupling 
\begin{align}\label{generic_vector_action_ghost}
&\frac{1}{16\pi} \int d^4x \sqrt{-g} [R-F_{\mu\nu}F^{\mu\nu} - 2m_X^2 X_\mu X^\mu] \nonumber \\
+&\frac{1}{16\pi} \int d^4x \sqrt{-g}\ \bar{\lambda}_\eta^2  f_\eta(\eta) \mathcal{L}_\chi[\chi, g_{\mu\nu}]]\ ,
\end{align}
with $\eta=F_{\mu\nu}F^{\mu\nu}/2$, and equation of motion
\begin{align}\label{generic_vector_eom_ghost}
\nabla_\rho  \left[(-\frac{\bar{\lambda}_\eta^2 }{2}\mathcal{L}_\chi f_\eta'(\eta) +1)\ F^{\rho \mu} \right]= m_X^2 X^\mu \ .
\end{align}
A choice like $f_\eta = 1-e^{\beta_\eta \eta/2}$ would provide ghost-based 
spontaneous  growth of the vector field. This theory also features higher derivatives in its
equations of motion, and its physical nature should be understood with
the same caveats as the scalar ghosts we discussed above.

\subsection{Spontaneous spinorization}
There are tachyon- and ghost-based spontaneous growth options for
scalars and vectors, but the derivative coupling is a necessity in some
cases which leads to ghostlike behavior. Let us try to continue the logic
of Sec.~\ref{sec:vectorization}, and investigate how a spinor can
spontaneously grow through the action
\begin{align}\label{generic_spinor_action}
\frac{1}{16\pi} \int d^4x \sqrt{-g} \bigg[ & R+\frac{1}{2} \left(\psib \gamma^\mu (\nabla_\mu \psi) 
- (\nabla_\mu \psib) \gamma^\mu \psi \right)
 \nonumber \\
&- m_\psi \psib \psi + \bar{\lambda}_\psi^2 f_\psi\ \mathcal{L}_\chi[\chi, g_{\mu\nu}] \bigg]
\end{align}
where $\psi$ is a four-component Dirac bispinor with mass $m_\psi$,
$\gamma^\mu$ are the curved-space Dirac gamma matrices that
satisfy $\gamma^{(\mu}\gamma^{\nu)}=g^{\mu\nu}$,  
and $\psib \equiv -i\psi^\dagger \gammab^0$ where
$\gammab^{(\mu}\gammab^{\nu)}=\eta^{\mu\nu}$ is the defining relationship of
the flat spacetime gamma matrices.
The spinor terms comprise what we saw in all of the theories in this 
study: the canonical action of the field that we want to grow. What
should $f_\psi$ be so that the equation of motion implies an instability
of $\psi=0$? The answer to this question is less transparent than its
counterparts for scalars or vectors. 

The terms that introduced the tachyonic
($\phi^2$, $X_\mu X^\mu$) and ghostlike instabilities ($K$, $\eta$) were all
quadratic in nature for scalars and vectors, and we could reverse their
signs to obtain an instability. The equation of motion for a spinor is
first order, and reversing the sign of the derivative term or the mass term
does not change its nature. Nevertheless, a solution has been found in the
context of generalizing the DEF theory to spinors, which  
utilizes the multicomponent nature of $\psi$~\cite{Ramazanoglu:2018hwk}.
The choice is $f_\psi = f_\psi(\mathcal{L}_\psi^{5,K})$ where
\begin{align}\label{fpsi}
2\mathcal{L}_\psi^{5,K}= \psib \gammab^5 \gamma^\mu (\nabla_\mu \psi) 
- (\nabla_\mu \psib) \gammab^5 \gamma^\mu \psi 
\end{align}
with 
\begin{align}\label{gamma5_curved}
\gamma^5 \equiv 
\frac{i}{4!} \epsilon_{\mu\nu\rho\sigma} \gamma^\mu\gamma^\nu\gamma^\rho\gamma^\sigma
=\frac{i}{4!} \tilde{\epsilon}_{abcd} \gammab^a\gammab^b\gammab^c\gammab^d = \gammab^5 \ ,
\end{align}
which is the flat-space gamma matrix.

We can see that $f_\psi(\mathcal{L}_\psi^{5,K})$ provides an instability 
through the equation of motion
\begin{align}\label{dirac_eqn_spinorization}
(\zeta_{\psi} \gammab^5 +\mathbb{I})\ \gamma^\mu \nabla_\mu \psi -  
[m_\psi-\frac{1}{2}(\nabla_\mu\zeta_\psi) \gammab^5 \gamma^\mu  ]\psi =0 \ ,
\end{align}
or after some basic manipulation
\begin{align}\label{dirac_eqn_spinorization2}
 \gamma^\mu \nabla_\mu \psi -  
\frac{\mathbb{I} - \zeta_{\psi} \gammab^5}{1-\zeta_{\psi}^2} 
[m_\psi-\frac{1}{2}(\nabla_\mu\zeta_\psi) \gammab^5 \gamma^\mu  ]  \psi = 0 \ .
\end{align}
where
\begin{align}\label{zeta_psi}
\zeta_\psi \equiv \bar{\lambda}_\psi^2\  \mathcal{L}_\chi f_\psi'(\mathcal{L}_\psi^{5,K})
\end{align}

Like the ghost-based spontaneous tensorization scenarios before,
spontaneous spinorization also brings higher derivative terms.
When $\mathcal{L}_\chi$ is a curvature term, the situation is
similar to that of ghost-based spontaneous scalarization and vectorization,
and we need to understand the theory as an effective one, unless there is 
a yet unidentified symmetry that cancels the higher derivatives.

When $\mathcal{L}_\chi$ is not a curvature term (e.g. Einstein-Maxwell-scalar
theories) there is no higher-derivative problem, but the nature of the spinor equation
seems to change drastically: $\zeta_\psi$ contains first derivatives of
$\psi$, and $\nabla_\mu\zeta_\psi$ seems to contain second derivatives of 
$\psi$. Hence, Eq.~\ref{dirac_eqn_spinorization2} is a second-order
partial differential equation in $\psi$ as opposed to the first-order
formulation in GR. However, this is not actually the case because
Eq.~\ref{dirac_eqn_spinorization2} implies
\begin{align}
\mathcal{L}_\psi^{5,K} = -m \frac{\zeta_\psi}{1-\zeta_\psi^2} \psib \psi \ ,
\end{align}
This, together with Eq.~\ref{zeta_psi}, is equivalent to
\begin{align}\label{Apsi}
\frac{\zeta_{\psi}}{\bar{\lambda}_\psi^2\  \mathcal{L}_\chi }=f_\psi' \left(-m \psib \psi\
\frac{ \zeta_{\psi}}{1-\zeta_{\psi}^2} \right)\ .
\end{align}
This implicit relationship can be used to express $\zeta_\psi$ as a  function of $\psib \psi$
and $\mathcal{L}_\chi $, but not their derivatives.  The spinor equation of motion remains
first order as a consequence.

Note that our initial aim was not a derivative coupling or a ghostlike
instability. We were merely trying to imitate the transition from
scalars to vectors, and finally to spinors, which naturally required a derivative
coupling. Hence, the effort to extend spontaneous growth in terms of types of
fields or instabilities, and the problems they bring with them, 
cannot always be separated. We should also add
that a spinor that grows spontaneously cannot satisfy the occupation
number rules of quantum mechanics, and hence it is
strictly a classical object~\cite{Ramazanoglu:2018hwk}.

\subsection{Higgs-based spontaneous growth}
Another mechanism to generate spontaneous growth can be 
obtained by noting that the coupling that leads to the spontaneous
growth in DEF and ESTGB theories acts as a mass
generation mechanism, albeit with the wrong sign. There is already
a well-known mass generation mechanism in physics, the Higgs
mechanism. Hence, the Higgs mechanism can also be used to incite
spontaneous growth by generating a mass-squared term with the
wrong sign. This recipe has been applied to extend the DEF theory
to a spontaneously growing vector field $X_\mu$ due to its coupling
to a complex Higgs scalar $\Phi$ as in the action~\cite{Ramazanoglu:2018tig}
\begin{align}\label{action_vt_higgs0}
\phantom{+}&\frac{1}{16\pi} \int d^4x \sqrt{-g} R 
- \frac{1}{16\pi} \int d^4x \sqrt{-g} F_{\mu\nu} F^{\mu\nu} \nonumber \\
-&\frac{1}{16\pi}\int d^4x \sqrt{-g} \big(2 \overline{D_\mu \Phi} D^\mu \Phi +2V(\Phib\Phi)\big) \nonumber \\
+& S_m \left[\psi_m, A_H^2 g_{\mu \nu} \right] \ ,
\end{align}
where $V(\Phib\Phi) = m_0^2(u^2 - \Phib\Phi)^2/(2u_0^2)$ is the Mexican-hat
potential of the Higgs field which provides its nonzero expectation
value, $D_\mu \Phi = (\nabla_\mu - i e X_\mu) \Phi$ is
the gauge-covariant derivative with coupling constant $e$, and
 $A_H= A_H(D^2)$ where $D^2 \equiv\overline{D_\mu \Phi} D^\mu \Phi$
and an overbar denotes complex conjugation. The argument of $A_H$ is
chosen in this way since the mass of $X_\mu$ is generated due to the
$\overline{\Phi} \Phi X_\mu X^\mu$ term inside this expression. This
means that an appropriate functional form of $A_H$ can invert the sign
of the effective vector field mass.

The vector field equation of motion is
\begin{align}\label{eom_higgs0}
\nabla^\nu F_{\nu\mu} &=(-8\pi \tilde{T} A_H^3 A_H'+1) 
(e^2 \Phib\Phi\ X_\mu + J^\Phi_\mu) \ ,
\end{align}
where $J^\Phi_{\mu} = ie (\Phib\nabla_\mu \Phi - \Phi \nabla_\mu \Phib )/2$.
The typical choice of
$A_H = e^{\beta_H D^2/2 }$ changes the sign of the effective mass of $X_\mu$,
and leads to spontaneous growth as explained in detail in Ref.~\cite{Ramazanoglu:2018tig}.

This mechanism is extended to more generic
couplings
\begin{align}\label{generic_higgs_action}
&\frac{1}{16\pi} \int d^4x \sqrt{-g} [R-F_{\mu\nu} F^{\mu\nu}
-2 \overline{D_\mu \Phi} D^\mu \Phi -2V(\Phib\Phi)] \nonumber \\
+&\frac{1}{16\pi} \int d^4x \sqrt{-g} f_H\ \mathcal{L}_\chi[\chi, g_{\mu\nu}]] \ ,
\end{align}
where $f_H = f_H(D^2)$ with the typical
example $f_H = 1-e^{\beta_H D^2/2}$.
The equation of motion is given by
\begin{align}\label{generic_higgs_eom}
\nabla^\nu F_{\nu\mu} =(-\frac{\mathcal{L}_\chi}{2} \widehat{f_H'(D^2)}+1) 
(e^2 \Phib\Phi\ X_\mu + J^\Phi_\mu)
\end{align}

The relationship between the Higgs mechanism and spontaneous
growth was first noted in the opposite direction, where spontaneous
growth is used instead of the potential $V(\Phib\Phi)$ to
provide a nonzero expectation value for $\Phi$, which is called
the gravitational Higgs mechanism~\cite{Coates:2016ktu}. 
It is possible to have a
second form of Higgs-based spontaneous growth through this
mechanism (see the discussion around
Eq.~26 in Ref.~\cite{Ramazanoglu:2018tig}).
Both the gravitational
Higgs mechanism and the Higgs-based spontaneous growth
mechanism inspired by it can also be adapted to generic couplings,
as in Eq.~\ref{generic_higgs_action}.

The Higgs mechanism can also provide the mass term
for non-Abelian gauge bosons, which is critical for their quantum
theory. Consequently, one can also spontaneously grow such fields
$W^a_\mu$ through the Higgs mechanism as well
in a theory with the action~\cite{Ramazanoglu:2018tig}
\begin{align}\label{generic_higgs_action2}
\frac{1}{16\pi} \int d^4x \sqrt{-g} \big[& R-F^{a\mu\nu} F^a_{\mu\nu}
-2(D_{\mu} \Phi)^\dagger D^\mu \Phi \nonumber \\
-&2V(\Phi^\dagger\Phi)  f_{YM}\ \mathcal{L}_\chi[\chi, g_{\mu\nu}] \big] \ ,
\end{align}
where
\begin{align}\label{YM0}
F^a_{\mu\nu} &= \nabla_\mu W^a_\nu - \nabla_\nu W^a_\mu + e f^{abc} W^b_\mu W^c_\nu \nonumber \\
D_\mu\Phi &=\nabla_\mu \Phi - ie W_\mu^b T^b \Phi \ ,
\end{align}
where $a,b,c$ label the generators of the Lie algebra of the gauge group ($T^a$),
$f^{abc}$ are the structure constants and $\dagger$ indicates Hermitian
conjugation. The natural choice for spontaneous growth is
$f_{YM}=1-e^{\beta_{YM} (D_{\mu} \Phi)^\dagger D^\mu \Phi/2}$.

The Higgs mechanism generates mass, and hence it can be considered
to be closely associated with a tachyonic instability. This is indeed the case
for the vector field in Eq.~\ref{generic_higgs_eom}, but the coupling terms
$\overline{D_\mu \Phi} D^\mu \Phi$ and $(D_{\mu} \Phi)^\dagger D^\mu \Phi$
also contain derivative expressions for the Higgs field itself. 
This means that even though Eq.~\ref{generic_higgs_eom} is free of the higher-derivative problems
we encountered for ghost-based scalarization and vectorization,
the equation of motion for $\Phi$ contains third-order derivatives of $g_{\mu\nu}$,
and the modified Einstein equation for $g_{\mu\nu}$ contains third-order derivatives of
$\Phi$. Our previous discussion about the physical meaning of derivative couplings
applies to all Higgs-based spontaneous growth theories as well.
\section{Discussion}\label{sec:discussion}
This study aimed to combine two approaches to gravity theories
with hairy compact object solutions: generalizing the scalar-dependent couplings
of the action from matter to more general terms, as in the ESTGB theory,
and generalizing the field that goes through spontaneous growth beyond the scalar,
as in spontaneous tensorization. The new family of extended theories
of spontaneous growth has the general action
\begin{align}\label{generic_all_action}
\frac{1}{16\pi} \int d^4x \sqrt{-g} [R+ \mathcal{L}_\xi
+f_\xi(\xi, \partial \xi) \mathcal{L}_\chi[\chi, g_{\mu\nu}]]\ ,
\end{align}
where $\xi$ is the field that spontaneously grows whose action is $\mathcal{L}_\xi$,
and  $\mathcal{L}_\chi$ is the Lagrangian that drives the spontaneous growth
which can be purely curvature (e.g., ESTGB), or may also depend on
other fields $\chi$ (e.g., Einstein-Maxwell-scalar).

The extended family has considerably richer phenomenology compared
to spontaneous scalarization. Black holes can also be scalarized, which is 
not possible in spontaneous tensorization. In
theories where the coupling is to a term that contains another field $\chi$,
the phenomenon goes beyond gravity, since such terms can be nonzero
even in flat spacetime and cause growth, as in
Einstein-Maxwell-scalar theories. The far-field behavior of a curvature term
such as the Gauss-Bonnet term is radically different from that of neutron star matter 
which only lives in a finite region of space. This means that the weak-field behavior
of extended spontaneous tensorization theories is also more varied compared to
spontaneous tensorization, which may provide further observable signatures.

We constructed these theories in a way that avoids higher derivative terms
of $\xi$ so that there are no obvious runaway instabilities. However,
when the coupling $f_\xi$ is to a curvature term, we saw that even a single derivative
of $\xi$ leads to more than two time derivatives
in the equations of motion.
Such theories can be examined under the umbrella of ``effective theories''.
An alternative, more conservative, approach is to disregard
any higher derivative theory, using our results as ``no-go'' theorems
to explore the limits of spontaneous growth phenomena. This latter 
path is also interesting since the original idea of spontaneous
tensorization applied to matter couplings does not suffer from
higher derivatives. This means, theories where the field couples to
different Lagrangians ($\mathcal{L}_\chi$) can be radically different.
Understanding the reasons for this can provide further insights into
nature of gravity in general.

When the coupling function $f_\xi$ does not have derivative dependence,
there is no higher-derivative problem, as in the extended spontaneous 
vectorization in Eq.~\ref{generic_vector_action}. However, this is not 
a complete guarantee of well-posedness. To start with,
there is already an instability in these theories to incite the spontaneous 
growth, which is known to be shut off in the DEF theory due to nonlinear
effects as the field grows~\cite{Ramazanoglu:2016kul}. This logic
can also be continued to extended scalar-tensor theories~\cite{Blazquez-Salcedo:2018jnn,Minamitsuji:2018xde,Silva:2018qhn},
but the mechanisms that regularize the instabilities have not been closely
investigated beyond these, and their effectiveness in the nonlinear regime
is not definitely known. Ghost-like instabilities need special
care since some of their solutions in DEF-like theories are known
to show unusual features in astrophysical systems even when
they do not lead to higher-derivative equations of motion~\cite{Ramazanoglu:2017yun}. 
It might be that there are more stringent requirements on $f_\xi$ than
the examples we mentioned here ($1-e^{\beta_\xi \dots}$), which
were based on the analysis of the linearized equations of motion
and common practices in the DEF theory~\cite{Ramazanoglu:2016kul}.

We should mention that the exact form of Eq.~\ref{generic_all_action} does not exhaust 
all cases of spontaneous growth. Many known examples are
closely related to this form such as axionic instabilities~\cite{Boskovic:2018lkj},
spontaneous growth of hidden vector fields~\cite{Annulli:2019fzq}, and
the extended massive gravity theories where the coupling between the
two metrics is a function of a dynamical scalar~\cite{Zhang:2017jze}.
However, one can look for even more general actions, for example,
by giving up our adopted condition that the action of the spontaneously growing
field is the canonical one in the Einstein frame. For the case of scalars, all possible action terms
that lead to second-order-in-time equations have been classified as
Horndeski~\cite{Horndeski1974} and beyond-Horndeski~\cite{PhysRevLett.114.211101}
theories. These may be starting points for more general forms of
spontaneous growth, and this strategy can also be repeated
for vectors as well.

The spontaneous growth mechanism in Eq.~\ref{generic_all_action}
can be adapted to other cases (such as more than one field)
in a relatively straightforward manner. However, it has
certain limits  which are similar to those of spontaneous tensorization
theories with matter coupling. For example, it does not seem to be
possible to find a theory that would feature spontaneously growing
spin-$2$ fields with local Lorentz symmetry, since any coupling between
two spin-$2$ fields other than that of de Rham-Gabadadze-Tolley
massive gravity would contain ghosts~\cite{deRham:2014zqa,Ramazanoglu:2017yun}.

The relationship between extended spontaneous tensorization
theories and their nonextended counterparts
is not merely a mathematical similarity, these theories share observable
signatures.
First, their deviation from GR is nonperturbativ (i.e. order-of-unity)
in the strong field regime.
This makes them a relatively easy target for observatories, and theoretical
predictions about one theory in the family can provide valuable
insight for the group as a whole. Second, the appeal of the DEF theory
was the fact that it avoided the constraints on the weak field that 
disfavored other scalar-tensor theories such as that of
Brans-Dicke theory~\cite{PhysRevLett.70.2220}. This feature is also shared by 
spontaneous tensorization, both extended or nonextended. 
Other modes of field growth such as dynamical scalarization in the 
DEF theory~\cite{2013PhRvD..87h1506B} are also likely to have
analogs in all extended spontaneous growth theories.
 
\acknowledgments
We thank Thomas Sotiriou for valuable discussions on the content of the
paper and on spontaneous growth phenomena in general.
The author is supported by Grant No.~117F295
of the Scientific and Technological Research Council
of Turkey (T{\"U}B{\.I}TAK). We would like to acknowledge networking and
travel support by the European Cooperation in Science and Technology
(COST) Action CA16104: GWverse.
\bibliography{/Users/fethimubin/research/papers/references_all}

\end{document}